\documentclass[apjl]{emulateapj}
\usepackage{comment}
\usepackage{color}
\usepackage{xspace}

\usepackage[perpage]{footmisc}

\slugcomment{Submitted to ApJL December 18, 2015. Accepted January 16, 2016.}
\shorttitle{Tidal Disruption Events Prefer Unusual Host Galaxies}
\shortauthors{French, Arcavi, \& Zabludoff}

\begin{document}

\title{Tidal Disruption Events Prefer Unusual Host Galaxies}

\author{
K. Decker French     \altaffilmark{1},
Iair Arcavi \altaffilmark{2},\altaffilmark{3}
Ann Zabludoff      \altaffilmark{1}}

\altaffiltext{1}{Steward Observatory, University of Arizona, 933 North Cherry Avenue, Tucson AZ 85721}
\altaffiltext{2}{Las Cumbres Observatory Global Telescope, 6740 Cortona Dr, Suite 102, Goleta, CA 93111}
\altaffiltext{3}{Kavli Institute for Theoretical Physics, University of California, Santa Barbara, CA 93106}

\begin{abstract}
Tidal Disruption Events (TDEs) are transient events observed when a star passes close enough to a supermassive black hole to be tidally destroyed. Many TDE candidates have been discovered in host galaxies whose spectra have weak or no line emission yet strong Balmer line absorption, indicating a period of intense star formation that has recently ended.  As such, TDE host galaxies fall into the rare class of quiescent Balmer-strong galaxies. Here, we quantify the fraction of galaxies in the Sloan Digital Sky Survey (SDSS) with spectral properties like those of TDE hosts, determining the extent to which TDEs are over-represented in such galaxies. Galaxies whose spectra have Balmer absorption H$\delta_{\rm A}$ $-$ $\sigma$(H$\delta_{\rm A}$) $>$ 4\,\AA\ (where $\sigma$(H$\delta_{\rm A}$) is the error in the Lick H$\delta_{\rm A}$ index) and H$\alpha$ emission EW $<$ $3$ \AA\ have had a strong starburst in the last $\sim$Gyr. They represent 0.2\% of the local galaxy population, yet host 3 of 8 (37.5\%) optical/UV-selected  TDE candidates. A broader cut, H$\delta_{\rm A} >$ 1.31\,\AA\ and H$\alpha$ EW $<$ $3$\, \AA, nets only 2.3\% of SDSS galaxies, but 6 of 8 (75\%) optical/UV  TDE hosts. Thus, quiescent Balmer-strong galaxies are over-represented among the TDE hosts by a factor of 33-190. The high-energy-selected TDE {\it Swift} J1644 also lies in a galaxy with strong Balmer lines and weak H$\alpha$ emission, implying a $>80\times$ enhancement in such hosts and providing an observational link between the $\gamma$/X-ray-bright and optical/UV-bright TDE classes.
\end{abstract}

 \keywords{galaxies: evolution, galaxies: nuclei}

%--------------------------------------------------------------------

---
\section{Introduction}

If a star passes close enough to a supermassive black hole that
the tidal forces overcome the self-gravity of the star, the star will be destroyed in a tidal disruption event \citep[TDE;][]{Hills1975}. Such events are expected to generate an observable flare \citep{Rees1988, Evans1989, Phinney1989} if the tidal radius is greater than
the Schwarzschild radius.

Real-time discoveries of TDE candidates have enabled extensive followup observations and classification. The first was \emph{Swift} J1644 \citep{Bloom2011, Burrows2011, Levan2011, Zauderer2011}, displaying non-thermal emission
in $\gamma$-rays, X-rays, and the radio. Two additional events had similar properties: \emph{Swift} J2058 \citep{BradleyCenko2012} and \emph{Swift} J1112 \citep{Brown2015}. Hereafter we refer to these events as ``high energy TDEs''.

In parallel, a different class of transients were also identified
as likely TDEs. The first was PS1-10jh \citep{Gezari2012}, which had thermal optical and UV emission but no observed X-rays. 
Since the discovery of PS1-10jh, eight objects with similar TDE spectral features have been found. \citet[hereafter A14]{Arcavi2014} discovered three in Palomar Transient Factory (PTF) data, PTF09axc, PTF09djl, and PTF09ge, grouping them together as a class with PS1-10jh, SDSS J0748 \citep[identified in the Sloan Digital Sky Survey (SDSS);][]{Wang2011}, and ASASSN-14ae \citep{Holoien2014}. Recently, three additional members of this class have been discovered: ASASSN-14li \citep{Holoien2015}, ASASSN-15oi \citep{Prentice2015}, and PTF15af (in the galaxy SDSS J084828.13+220333.4; Blagorodnova et al., in prep). This optical/UV-selected class of transients all display hot blackbody ($\sim\textrm{few}\cdot10^{4}$K) emission and several-month-long smooth light curves peaking at an absolute optical magnitude of $\sim-20$. They are all located in the centers of their host galaxies. Their clear broad H and/or He emission lines (A14) cleanly distinguish them from other transient events. Hereafter we refer to these events as ``optical/UV TDEs''.

Curiously, the host galaxy spectra of these eight\footnote{As of submission, the TDE ASASSN-15oi was still ongoing, and no uncontaminated optical spectrum of the host galaxy was available.} events show Balmer line absorption, and all but SDSS J0748 have weak or no emission lines. This combination of spectral features indicates low levels of current star formation, yet substantial star formation in the last $\sim$ Gyr, long enough ago for the ionizing O and B stars to have evolved away, but recently enough for A stars to dominate the stellar light. Galaxies with these spectral features are called Balmer-strong, H$\delta$-strong, E+A, k+a, or a+k galaxies, depending on the strength of the Balmer absorption and the authors' preference \citep{Dressler1983,Couch1987,Dressler1999}. Many are consistent with a post-starburst star formation history and galaxy-galaxy merger origin, and are likely in transition between star-forming spirals and passive early-type galaxies (\citealp{Zabludoff1996}; \citealp[Y.][]{Yang2004,Yang2008}).  

Similarly, the host galaxy of the high energy TDE candidate {\it Swift} J1644 was reported to have significant Balmer absorption with a low current star formation rate (SFR) predicted from its H$\alpha$ flux \citep{Levan2011}. \citet{Yoon2015} find a young $<1$ Gyr stellar population, suggesting a recent starburst. The two other known high energy TDE candidates do not yet have host galaxy spectra covering the full Balmer series.

The clear preference of optical/UV TDE candidates and at least one high energy TDE candidate for rare Balmer-strong galaxies has important implications for the mechanisms driving TDE rates. Here we examine the statistics of galaxies like the TDE candidate hosts to quantify the TDE-rate enhancement in such environments.

%%%%%%%%%%%%%%%%%% figures 
\begin{figure*}
\includegraphics[width=1\textwidth]{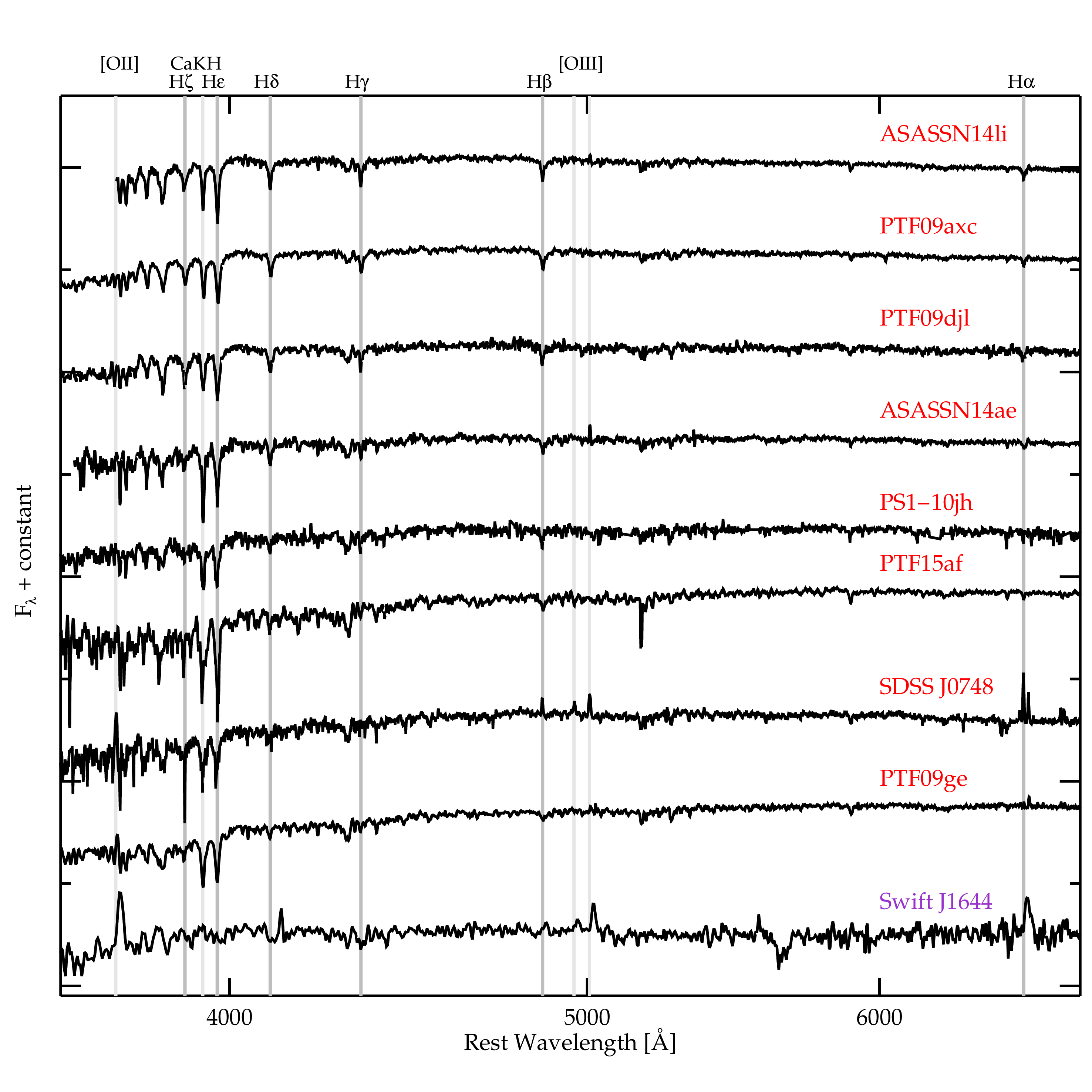}
\caption{Spectra of the eight optical/UV  TDE host galaxies, in order of decreasing strength of their H$\delta_{\rm A}$ index. Also shown is the lower-resolution host galaxy spectrum of the high energy TDE {\it Swift} J1644. Strong Balmer absorption, Ca II H+H$\epsilon$ absorption, and a lack of strong emission lines are characteristic of post-starburst galaxies. Both SDSS J0748 and {\it Swift} J1644 were selected differently from the rest of the sample, although the optical spectrum of the TDE itself in SDSS J0748 appears similar to the other optical/UV TDEs.}
\label{fig:spec}
\end{figure*}

\begin{figure*}
\includegraphics[width=1.0\textwidth]{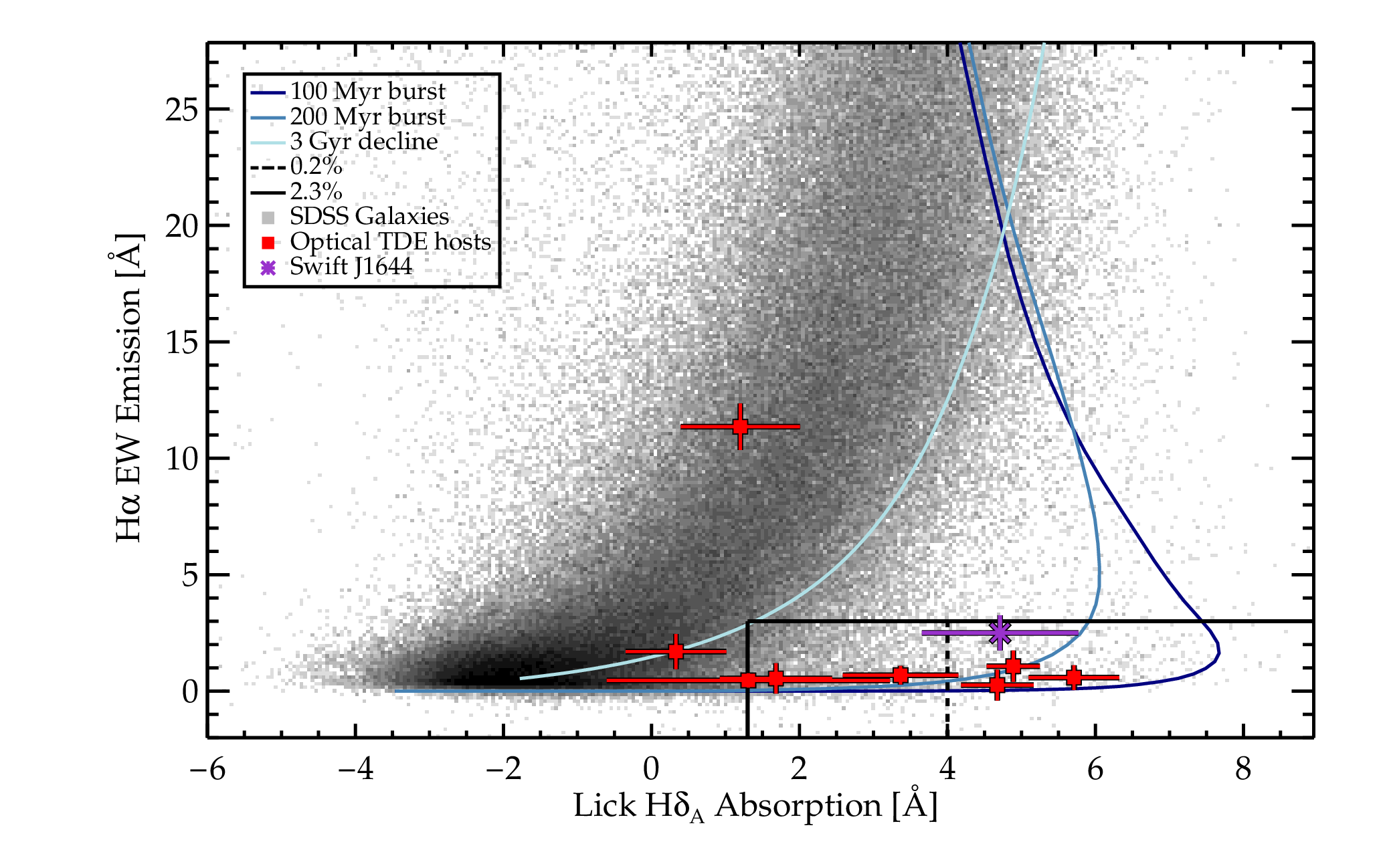}
\caption{Spectral characteristics of SDSS galaxies (grey) and TDE candidate host galaxies (colored points): H$\alpha$ EW emission (current star formation) versus H$\delta_{\rm A}$ absorption (from A stars, indicating star formation within the past $\sim$Gyr). The SDSS galaxies populate the ``red sequence'' (low H$\alpha$ EW, low H$\delta_{\rm A}$) and ``blue cloud'' (extending up to higher H$\alpha$ EW at moderate H$\delta_{\rm A}$). Many TDE hosts lie within the quiescent Balmer-strong galaxy ``spur'' extending to high H$\delta_{\rm A}$ at low H$\alpha$ EW. 
Two cuts along the spur are shown: H$\alpha$ EW $<$3 \AA\ with H$\delta_{\rm A}$ $-$ $\sigma$(H$\delta_{\rm A}$) $>$ 4\,\AA\ (dashed boundary) and H$\delta_{\rm A} > 1.31$\, \AA\ (solid boundary). These regions include only 0.2\% and 2.3\% of the SDSS galaxies, yet encompass 38\% and 75\% of the optical/UV TDE host galaxies, respectively. Three example star formation history tracks are shown. Short duration starbursts (dark and medium blue) on top of an existing old stellar population will pass through the strongest H$\delta_{\rm A}$ region once the starburst ends, evolving through the moderately strong H$\delta_{\rm A}$ region at later times. A gradually declining star formation history (light blue) cannot pass through the strictest H$\delta_{\rm A}$ cut. TDE host galaxies with the highest H$\delta_{\rm A}$ absorption thus have likely experienced a recent starburst. Galaxies with H$\delta_{\rm A} = 1.3$ to $4$\AA\ have a range of possible star formation histories (see text), but have still experienced a recent decline in their star formation. The TDE hosts SDSS J0748 and PTF09ge do not lie in the spur, but among the star-forming and early-type populations, respectively. The high energy TDE candidate {\it Swift} J1644 (purple) has strong H$\delta_{\rm A}$ absorption (its errors place it just outside our strictest cut). Even if {\it Swift} J1644 turns out to be the only one of the three known high energy TDEs with a host that lies in this region, high energy TDE rates will be over-represented in quiescent Balmer-strong galaxies by $>80\times$.}
\label{fig:ha_hd}
\end{figure*}

%%%%%%%%%%%%%%%%%%%%%%%

\section{Spectral Properties of TDE Host Galaxies}
\subsection{Quiescent Balmer-Strong Classification}
We wish to determine the Balmer stellar absorption line and nebular emission line properties of TDE hosts to quantify the incidence of galaxies like them. Balmer absorption is generally characterized using the H$\delta$ line, due to its low emission filling and smooth nearby continuum regions. \citet{Dressler1999} identify quiescent galaxies with H$\delta$ EW$>3$\AA\ as ``k+a'' or ``Balmer-strong'' galaxies. Here, we parameterize Balmer absorption using the Lick H$\delta_{\rm A}$ index, which is optimized for the stellar absorption from A stars \citep{Worthey1997}. The H$\delta_{\rm A}$ measure differs slightly from H$\delta$ EW: a cut on H$\delta_{\rm A}>4$\AA\ is equivalent to H$\delta$ EW $>3$\AA. For our strictest cut, we require H$\delta_{\rm A}$ $-$ $\sigma$(H$\delta_{\rm A}$) $>$ 4\,\AA, where $\sigma$(H$\delta_{\rm A}$) is the H$\delta_{\rm A}$ index measurement error included to eliminate spurious objects with large $\sigma$(H$\delta_{\rm A}$). We do not correct H$\delta_{\rm A}$ for emission line filling, as the correction is smaller than the measurement error for galaxies with both strong H$\delta$ and weak H$\alpha$ emission. 

We select for galaxies with little on-going star formation by requiring H$\alpha$ EW $<$ 3\,\AA\ in emission in the rest frame, corresponding to a specific SFR $\lesssim 1\times10^{-11}$ yr$^{-1}$, well below the main sequence of star-forming galaxies \citep[e.g.,][]{Elbaz2011}. We correct H$\alpha$ EW for stellar absorption, which is significant for quiescent Balmer-strong galaxies.% Due to the strong Balmer absorption, H$\alpha$ is seen in absorption for many of these galaxies, and we must correct for the stellar absorption in order to isolate the residual H$\alpha$ emission indicative of current star formation. 

\subsection{Data}
Optical spectroscopy was obtained for the eight host galaxies of the optical/UV TDE candidates from A14, C. \citet{Yang2013}, and the SDSS, and for the host of high energy TDE candidate {\it Swift} J1644 from \citet{Levan2011}. We plot these host galaxy spectra in Figure \ref{fig:spec}, describing their features in Table 1.

For each TDE candidate host galaxy, we calculate H$\delta_{\rm A}$ as described above, and H$\alpha$ EW using two pseudo-continuum bands (Blue: [6492.8-6532.8\AA]; Red: [6592.8-6632.8\AA]). We use stellar population models from \citet{Conroy2009} and \citet{Conroy2010}  to fit archival UV and optical photometry (from SDSS, {\it GALEX}, and {\it Swift} UVOT) together with the optical spectral features. We model the star formation history as an old population plus a recent, exponentially-declining burst of star formation. The age since the recent burst, mass produced and duration of the recent burst, and dust extinction are free parameters (French et al. in prep). From these models, we determine the stellar absorption correction to the H$\alpha$ EW, which ranges from 1.5-2.5\AA.

To quantify the rarity of these TDE hosts, we compare their spectral features to the SDSS main galaxy spectroscopic sample \citep{Strauss2002}, defining a parent sample from DR10 \citep{Aihara2011}. To prevent severe aperture bias, we exclude galaxies with $z <0.01$ to eliminate those that are very large on the sky relative to the 3\arcsec \ diameter of the SDSS fibers.  We also exclude galaxies with unreliable \footnote{We require \texttt{h\_alpha\_eqw\_err > -1}.} H$\alpha$ equivalent widths (EW) or median signal-to-noise values of less than 10 per pixel over the whole spectrum. Our final parent sample is composed of 591,736 galaxies. We use the H$\alpha$ EWs (corrected for stellar absorption) and Lick H$\delta_{\rm A}$ values from the MPA-JHU catalogs \citep{Brinchmann2004}. In the three cases where the TDE host galaxy spectrum is from the SDSS, our H$\delta_{\rm A}$ and corrected H$\alpha$ EW measurements are consistent with the MPA-JHU catalog values.

\begin{table*}
\centering
\caption{TDE Host Properties}
\label{table:data}
\begin{tabular}{l r r r r r l}
\hline
\hline
\multicolumn{1}{l}{TDE Host} &
\multicolumn{1}{c}{H$\alpha$ EW\footnotemark[1]} &
\multicolumn{1}{c}{H$\delta_{\rm A}$} &
\multicolumn{1}{c}{z} &
\multicolumn{1}{c}{$M_r$\footnotemark[2]} &
\multicolumn{1}{c}{Slit Width} &
\multicolumn{1}{l}{Data Source} \\

\multicolumn{1}{c}{} &
\multicolumn{1}{c}{[\AA]} &
\multicolumn{1}{c}{[\AA]} &
\multicolumn{1}{c}{} &
\multicolumn{1}{c}{mag} &
\multicolumn{1}{c}{[arcsec (kpc)]} &
\multicolumn{1}{c}{} \\
\hline
SDSS J0748 & -11.36$\pm$1.00 & 1.20$\pm$0.81 &   0.0615 &  -20.13$\pm0.02$ &    1.0 (1.2) & C. \citet{Yang2013}\\
ASASSN14ae & -0.68$\pm$0.40 & 3.37$\pm$0.79 &  0.0436 &  -19.75$\pm0.02$ &    3.0 (2.6) & SDSS \\
ASASSN14li & -0.59$\pm$0.53 & 5.71$\pm$0.61 &   0.02058 &   -19.20$\pm0.02$ &   3.0 (1.3) & SDSS  \\
PTF09axc & -1.07$\pm$0.67 & 4.89$\pm$0.36 &   0.1146 &  -20.55$\pm0.02$  &   1.0 (2.1) & A14 \\
PTF09djl & -0.26$\pm$0.66 & 4.67$\pm$0.49 &    0.184 &   -20.02$\pm0.03$ &   1.0 (3.1) & A14\\
PTF09ge & -1.70$\pm$0.75 & 0.33$\pm$0.68 &    0.064 &   -20.23$\pm0.02$ &  0.7 (0.9) & A14\\
PS1-10jh & -0.54$\pm$0.65 & 1.68$\pm$0.76 &    0.1696 &   -18.48$\pm0.05$ &   1.0 (2.9) & A14\\
PTF15af & -1.65$\pm$0.30 & 1.31$\pm$1.91 &      0.0790 &   -20.20$\pm0.02$  &  3.0 (4.5) & SDSS \\
\hline
{\it Swift} J1644 & -2.50$\pm$0.76 & 4.71$\pm$1.06 & 0.3534  &  -18.44$\pm0.1$ & 1.0 (5.0) & \citet{Levan2011} \\
\hline
\multicolumn{7}{l}{%
  \begin{minipage}{12cm}%
    Notes: \footnotetext[1]{Negative values indicate emission. H$\alpha$ EW values are corrected for stellar absorption.} \footnotetext[2]{Absolute magnitudes ($r$ band, no extinction correction) from SDSS ({\tt model\_mag}) for optical/UV TDE hosts, and from \citet{Levan2011} for {\it Swift} J1644. We assume $H_0=70$ km s$^{-1}$ Mpc$^{-1}$, $\Omega_\Lambda=0.7$, and $\Omega_m=0.3$.}

  \end{minipage}%
}\\
\end{tabular}
\end{table*}

\section{Likelihood of Host Galaxy Properties}
The scarcity of quiescent Balmer-strong galaxies makes it unlikely to see many TDE host galaxies in this class, were TDEs to occur in all types of galaxies. Our strictest selection cut, H$\delta_{\rm A}$ $-$ $\sigma$(H$\delta_{\rm A}$) $>$ 4\,\AA\ and H$\alpha$ EW $<$ 3\,\AA\ (emission), on the SDSS produces a sub-sample of 1207 galaxies (0.20\%). Of the optical/UV TDE host galaxies studied here, three (ASASSN14li, PTF09axc, PTF09djl) of the eight satisfy these cuts. Quiescent Balmer-strong galaxies are thus over-represented in the TDE host galaxy sample by a factor of 190$^{+115}_{-100}$ times.\footnote{Errors represent 1$\sigma$ binomial confidence intervals calculated using the small numbers tables from \citet{Gehrels1986}.}

The significance of these cuts is illustrated in Figure \ref{fig:ha_hd}. SDSS galaxies fall into the ``blue cloud,'' with H$\alpha$ emission and moderate H$\delta_{\rm A}$ absorption, and the ``red sequence,'' with little-to-no H$\alpha$ emission and low H$\delta_{\rm A}$. Quiescent Balmer-strong galaxies lie in the ``spur'' to the lower right, with weak H$\alpha$ emission and strong H$\delta_{\rm A}$ absorption. Most of the TDE host galaxies lie along this spur.

The example model tracks in Figure \ref{fig:ha_hd} show how galaxies with different star formation histories evolve. Two tracks follow a recent burst of star formation (an exponentially-declining model of short duration; $\tau=$ 100 or 200 Myr) on top of an existing old stellar population. We again use the stellar population models from Conroy et al. (2009, 2010), assuming that the recent burst created 10\% of the stellar mass of the galaxy. A shorter burst with the same mass fraction produces higher H$\delta_{\rm A}$ absorption. Star formation declining over $\tau>200$ Myr does not generate high enough H$\delta_{\rm A}$ to meet our strictest criterion. Thus, the three TDE host galaxies satisfying the strictest H$\delta_{\rm A}$ cut have probably experienced a recent, strong, brief period of star formation, i.e., a true burst.

The H$\delta_{\rm A}$ requirement can be relaxed to $>$ 1.31\,\AA\ to encompass the three other optical/UV TDE host galaxies that also lie in the spur: ASASSN14ae, PTF15af, and PS1-10jh. This cut includes 13749 SDSS galaxies, or 2.3\%, but six of the eight known TDE hosts. Thus, quiescent galaxies with at least moderately strong Balmer absorption are over-represented in the TDE host galaxy sample by a factor of 33$^{+7}_{-11}$. This lower H$\delta_{\rm A}$ cut allows several possible star formation histories, including star formation that has declined over several Gyr, starbursts that have ended and evolved past the strong H$\delta_{\rm A}$ region, and constant star formation with a sudden truncation \citep[e.g.,][]{Shioya2004}. Therefore, it is less straightforward to interpret the recent histories of the three lower H$\delta_{\rm A}$ galaxies; all have recently-ended periods of star formation, but may not have experienced a strong starburst. 

The two remaining TDE candidate host galaxies, SDSS J0748 and PTF09ge, do not lie in the spur of Figure \ref{fig:ha_hd}. The TDE in SDSS J0748 was discovered in a search for high-ionization emission lines \citep{Wang2011}, not in a transient survey, a selection that might contribute to the distinct spectral properties of its host, which lies in the blue cloud of star-forming galaxies in Figure \ref{fig:ha_hd}. In contrast, PTF09ge, found in the same manner as the other PTF A14 TDE candidates, has a host galaxy that lies closer to the red sequence of early-type galaxies.

Two of the optical/UV TDE host galaxies (PTF09djl and PS1-10jh) lie in the SDSS footprint, and have absolute magnitudes comparable to the other six host galaxies, but are too faint for the SDSS spectroscopic sample due to their slightly higher redshifts ($z=0.1696, 0.184$). No significant galaxy evolution from $z\sim0.2$ to $0.01$ is expected \citep{Snyder2011}. Therefore, the quiescent Balmer-strong galaxy fraction should remain the same. To check this, we cut the parent sample to bound the redshifts of these two TDE hosts. The resulting TDE rate enhancements are within the 1$\sigma$ errors of those quoted above.

The high energy TDE host galaxy {\it Swift} J1644 has some H$\alpha$ emission (Figure \ref{fig:spec}), but it is weak enough to meet our criteria for being quiescent. The host galaxy spectrum is noisy around H$\delta_{\rm A}$, and thus does not meet our strictest Balmer-strong cut of H$\delta_{\rm A}$ $-$ $\sigma$(H$\delta_{\rm A}$) $>$ 4\,\AA. However, stronger H$\beta$ (EW $=8.2\pm1.1$\AA) and H$\gamma$ (EW $=11.7\pm2.4$\AA) absorption are reported by \citet{Levan2011}. Because this galaxy is at a higher redshift ($z=0.3534$) than the optical/UV TDEs, we constrain the SDSS comparison sample to $z>0.3$ in this case. A {\it Swift} J1644-consistent cut on H$\delta_{\rm A}$ $-$ $\sigma$(H$\delta_{\rm A}$) $>$ 3.65\, \AA\ and H$\alpha$ EW $<$ 3 \AA\ yields 10 out of 2287 galaxies, or 0.4\%. Although we do not have full optical spectra for the other two high energy TDEs, it is significant that at least 1/3 of that sample lies within only 0.4\% of the parameter space.

\section{Discussion}
\label{disc}

\subsection{Preference of TDEs for Quiescent Balmer-Strong Galaxies}
\label{mech}

There are several unique characteristics of quiescent Balmer-strong galaxies that might act to boost the TDE rates. Many such galaxies have had a recent galaxy-galaxy merger, which might in turn lead to 1) a black hole binary, and/or 2) perturbed stellar orbits that pass closer to the black hole. A recent starburst associated with the merger could produce 3) a large A star population now evolving into more easily-disrupted giants, and/or 4) A stars concentrated in the galaxy's core. It is also not yet clear whether the high incidence of LINER-like emission in post-starburst galaxies (\citealp{Yan2006}; \citealp[Y.][]{Yang2006}) is related to TDEs, perhaps via residual gas in the nucleus.

In post-merger systems, a subsequent merger of the central black holes is expected, with minor mergers producing unequal black hole binaries. \citet{Chen2009,Chen2011} suggest that the TDE rate will be higher for unequal mass black hole binaries, as stars around one black hole are scattered into the other. \citet{Li2015} predict an increased TDE rate for a few tens of Myr once two equal mass black holes form a bound binary. The range in H$\delta_{\rm A}$ absorption in our sample suggests a range in merger ratios and/or a range of post-merger ages. In future work, it will be interesting to narrow down what TDE host physical properties correlate best with TDE rate enhancement.

A recent merger could also create an asymmetric central potential, which would alter stellar orbits near the nucleus, allowing for more centrophillic orbits than a spherically symmetric potential \citep{Magorrian1999}. Centrophillic orbits would bring more stars through the region where they could be tidally disrupted, increasing the TDE rate.

In galaxies that have experienced a starburst ending in the past $\sim$Gyr, the first A stars will be evolving off the main sequence. During their giant phase, these stars will be more easily disrupted due to their large envelopes \citep{MacLeod2012}. These evolving giants may also be susceptible to multiple epochs of mass loss to the black hole (and thus produce multiple TDEs), although such events might have different signatures than the TDEs here \citep{MacLeod2013}. It is not yet possible to tell if these TDEs are partial or full disruptions, and of which kinds of stars.

Post-starburst galaxies are known to have centrally concentrated distributions of A stars \citep[Y.][]{Yang2004,Yang2008} as a result of gas driven to the center during the merger. The increased stellar density in the core could serve to increase the TDE rate \citep{Stone2015}, and could act in combination with the other effects discussed here.

Is there an observational bias favoring the detection of TDEs in quiescent Balmer-strong galaxies? We consider two possibilities---that somehow flares are more readily seen or broad H/He spectral features more easily distinguished in these hosts---and discount them. For example, while detecting a central transient would be easier in a dust-poor or bulgeless host, quiescent Balmer-strong galaxies have global extinctions between star-forming and early-type galaxies \citep[e.g.,][]{Wild2009} and are bulge dominated with high S\'ersic indices \citep{Yang2008}. Furthermore, their bright spectral continuum, arising from their recent star formation, makes broad emission line detection harder, not easier.

Several of the TDE rate enhancement mechanisms described above should also operate in {\it currently} starbursting galaxies. There, we might expect to see a bias against the detection of TDEs, as starbursting/ULIRG galaxies will have significant dust attenuation \citep[e.g.,][]{Casey2014}. If TDEs occur mainly via black hole merger or starburst-related mechanisms, their lack of observability in starbursting galaxies must be considered in calculating the true rate of TDEs.

\subsection{Implications for the TDE Rate}
\label{tdechoice}

TDE rates derived from observations \citep[e.g.,][]{vanvelzen2014,Holoien2015} are $\sim 10^{-5}$ yr$^{-1}$ per galaxy, in tension with higher theoretical predictions of a few $\times 10^{-4}$ yr$^{-1}$ per galaxy \citep{Stone2015}.

Our results suggest that this rate is 33$^{+7}_{-11}$ to 190$^{+115}_{-100}$ times higher in quiescent galaxies with moderately strong and strong H$\delta_{\rm A}$ (Figure \ref{fig:ha_hd}). Extrapolating the observed TDE rate, this implies a TDE rate of $2-4\times 10^{-4}$ and $1-3\times 10^{-3}$ yr$^{-1}$ per galaxy, respectively. These TDE rate enhancements imply a lower TDE rate for normal star-forming and early-type galaxies, of $1-5\times 10^{-6}$ yr$^{-1}$ per galaxy.

These rate estimates do not include the high energy TDE {\it Swift} J1644. However, its host properties are so unusual that, even without knowledge of the two other TDE hosts, high energy TDE rates appear to be boosted by $>80\times$ in quiescent Balmer-strong hosts.

We have focused on a homogeneous sample of TDEs whose broad H and/or He emission lines distingish them from other types of transients. In doing so, we have excluded three other optical/UV TDE candidates with spectral identifications, but without clearly broadened emission lines. The first, TDE2 \citep{vanVelzen2011}, is the only borderline case, as its spectrum suggests a possible broad H feature, but with lower width and signal-to-noise. Its host's SDSS spectrum passes our H$\delta_{\rm A}>1.3$\AA\ cut, but including it in our analysis only changes the over-representation of TDEs in such hosts from 33 to 34$\times$. The last two, PTF10iya \citep{Cenko2012} and PS1-11af \citep{Chornock2014}, lie in a star-forming and quiescent galaxy, respectively, each with some evidence for Balmer absorption. Including these events would not diminish the over-representation of Balmer-strong galaxies as TDE hosts.

\section{Conclusions}
We demonstrate the preference of tidal disruption events (TDEs) to occur in quiescent galaxies with strong Balmer line absorption.  Quiescent galaxies with the strongest Balmer absorption, H$\delta_{\rm A}$ $-$ $\sigma$(H$\delta_{\rm A}$) $>$ 4\,\AA, make up only 0.2\% of local galaxies, yet host 3 of 8 optical/UV TDE candidates.  A softer cut, H$\delta_{\rm A} > 1.31$\AA, includes only
2.3\% of local galaxies, but 6 of 8 optical/UV TDE host galaxies. 
The optical/UV TDE rates are thus enhanced by 190$^{+115}_{-100}\times$ in the strongest H$\delta_{\rm A}$ galaxies and by 33$^{+7}_{-11}\times$ in galaxies with H$\delta_{\rm A}>1.31$\AA. Because of this preference, the corresponding rates of optical/UV TDEs are $1-3\times 10^{-3}$ yr$^{-1}$ per galaxy, and $2-4\times 10^{-4}$ yr$^{-1}$ per galaxy, respectively. As a result, we predict a lower optical/UV rate in normal star-forming and early-type galaxies, of $1-5\times 10^{-6}$ yr$^{-1}$ per galaxy. Even the one high energy TDE with a full measured optical spectrum, {\it Swift} J1644, lies in a galaxy with strong Balmer absorption and weak nebular line emission, which implies a $>80\times$ enhancement in such hosts, and a link between the optical/UV and high energy TDE classes.

Why do TDEs prefer quiescent, Balmer-strong hosts? This type of galaxy has several properties that make it special and may suggest an answer. Many have had a recent galaxy-galaxy merger \citep{Zabludoff1996}, increasing the possibility of a black hole binary, perturbed stellar orbits, a spatially-concentrated population of A stars, and/or an evolved population of easily-disrupted A giants. The high incidence of LINER-like emission (\citealp{Yan2006}; \citealp[Y.][]{Yang2006}) may also play a role. In future work, we will explore these connections.

\smallskip
We thank A. Levan and A. de Ugarte Postigo for providing the digital version of their {\it Swift} J1644 host galaxy spectrum. We thank Yujin Yang for his contributions to the stellar population fitting routines. KDF acknowledges support from NSF grant DGE-1143953, P.E.O., and the ARCS Phoenix Chapter and Burton Family.  IA is grateful for support from the Karp Discovery Award. AIZ and IA thank the Center for Cosmology and Particle Physics at NYU for its support. AIZ acknowledges funding from NSF grant AST-0908280 and NASA grants ADP-NNX10AD476/ADP-NNX10AE88G. AIZ thanks the John Simon Guggenheim Foundation for its support.

Funding for SDSS-III has been provided by the Alfred P. Sloan Foundation, the Participating Institutions, the National Science Foundation, and the U.S. Department of Energy Office of Science. The SDSS-III web site is http://www.sdss3.org/.

\bibliographystyle{apj}
%\bibliography{tde_refs.bib}

\end{document}